\documentclass[twocolumn,showpacs,aps,prb]{revtex4-1}
\usepackage{bm}
\usepackage{amssymb,amsmath}
\usepackage[dvips]{graphicx}

\begin{document}

\title{Quantum phase transition of the sub-Ohmic rotor model}

\author{Manal Al-Ali}

\author{Thomas Vojta}
\affiliation{Department of Physics, Missouri University of Science and Technology, Rolla, MO 65409, USA}

\begin{abstract}
We investigate the behavior of an $N$-component quantum rotor coupled to a bosonic
dissipative bath having a sub-Ohmic spectral density $J(\omega) \propto \omega^s$ with $s<1$.
With increasing dissipation strength, this system undergoes a quantum phase transition
from a delocalized phase to a localized phase. We determine the exact critical behavior
of this transition in the large-$N$ limit. For $1>s>1/2$, we find nontrivial critical
behavior corresponding to an interacting renormalization group fixed point while we find mean-field behavior
for $s<1/2$. The results agree with those of the corresponding long-range interacting classical
model. The quantum-to-classical mapping is therefore valid for the sub-Ohmic rotor
model.

\end{abstract}

\date{\today}
\pacs{05.30.Rt,05.30.Jp}

\maketitle

\section{Introduction}

Quantum phase transitions are abrupt changes in the ground state properties of a quantum
many-particle system that occur when a non-thermal control parameter is varied.\cite{Sachdev_book99}
In analogy to
thermal phase transitions, they can be classified as either first-order or continuous transitions.
Continuous quantum phase transitions, also called quantum-critical points, are characterized by
large-scale temporal and spatial fluctuations that lead to unconventional behavior in systems
ranging from strongly correlated electron materials to ultracold quantum gases
(for reviews see, e.g., Refs.\
\onlinecite{SGCS97,Vojta_review00,VojtaM03,ColemanSchofield05,GegenwartSiSteglich08,Sachdev08}).

Impurity quantum phase transitions\cite{VojtaM06} are an interesting class of quantum phase
transitions at which
only the degrees of freedom of a finite-size (zero-dimensional) subsystem become critical at the
transition point. The rest of the system (the ``bath'') does not undergo a transition.
Impurity quantum phase transitions can occur, e.g., in systems composed of a single quantum spin
coupled to an infinite fermionic or bosonic bath. Fermionic examples include the anisotropic Kondo
model\cite{AndersonYuvalHamann70} and the pseudogap Kondo model.\cite{WithoffFradkin90}

The prototypical system involving a bosonic bath is the dissipative two-state
system,\cite{LCDFGZ87,Weiss_book93} also called the spin-boson model,
which describes a two-level system coupled to a single dissipative bath of harmonic oscillators.
Its ground-state phase diagram depends on the behavior of the bath spectral density $J(\omega)$
for small frequencies $\omega$.
Power-law spectra $J(\omega) \propto \omega^s$ are of particular interest.
In the super-Ohmic case ($s>1)$, the system is in the
delocalized (disordered) phase for any dissipation strength. In contrast, for sub-Ohmic
dissipation ($0<s<1$), there is a continuous quantum phase transition from a delocalized phase
at weak dissipation to a localized (ordered) phase at strong dissipation.\cite{BullaTongVojta03}
In the marginal Ohmic case ($s=1$), a quantum phase transition exists, too, but it is of Kosterlitz-Thouless
type.\cite{LCDFGZ87,Weiss_book93}

The sub-Ohmic spin-boson model has recently attracted considerable attention
in the context of the so-called quantum-to-classical mapping. This concept relates the critical
behavior of a quantum phase transition in $d$ space dimensions to that of a classical transition in
$d+1$ dimensions. The mapping is usually established by comparing the order-parameter field theories
of the transitions: Imaginary time in the quantum problem plays the role of the extra dimension
in the corresponding classical system. In the case of the spin-boson model, the classical
counterpart is a one-dimensional Ising model with long-range interactions that decay as
$1/r^{1+s}$ for large distances $r$.
In recent years, the applicability of the quantum-to-classical
mapping to the sub-Ohmic spin-boson model has been controversially discussed after numerical
renormalization group results\cite{VojtaMTongBulla05} suggested that its critical behavior for $s<1/2$
deviates from that of the corresponding Ising model. While there is now strong
evidence\cite{WRVB09,AlvermannFehske09,VBGA10,TongHou10} that this conclusion is incorrect and that the
quantum-to-classical mapping is actually valid, the issue appears to be still not fully
settled.\cite{Kirchner10} Moreover, possible failures of the
quantum-to-classical
mapping have also been reported for other impurity models
with both Ising\cite{GlossopIngersent05,KirchnerSiIngersent09,ChengGlossopIngersent09}
and higher\cite{SachdevBuragohainVojta99,ZKSG04} symmetries;
and the precise conditions under which it is supposed to hold are not resolved.

In the present paper, we therefore investigate the large-$N$ limit of the sub-Ohmic quantum rotor model.
Analogously to the spin-boson model, this system undergoes a quantum phase transition with increasing
dissipation strength from a delocalized phase to a localized phase.\cite{CGLLS02,DrewesArovasRenn03}
We exactly solve the critical properties of this transition. Our analysis yields nontrivial critical behavior
corresponding to an interacting renormalization group fixed point for $1>s>1/2$,
while we find mean-field behavior for $s<1/2$. All critical exponents agree with those of
the corresponding long-range interacting classical model,\cite{Joyce66}
implying that the quantum-to-classical mapping is valid.

Our paper is organized as follows.  We define the sub-Ohmic rotor model in Sec.\ \ref{sec:model}.
In Sec.\ \ref{sec:constraint}, we derive its partition function; and we solve the self-consistent large-$N$ constraint
at zero and finite temperatures as well as with and without an external field.
Section \ref{sec:observables} is devoted to a discussion of observables and the resulting critical
behavior. We conclude in Sec.\ \ref{sec:conclusions}.

\section{Sub-Ohmic rotor model}
\label{sec:model}

A quantum rotor can be understood as a point moving on an $N$-dimensional hypersphere
of radius $N^{1/2}$. It can be represented by an $N$-component vector $\mathbf{S}$
satisfying $\mathbf{S}^2=N$. The rotor has a momentum $\mathbf{P}$; the position and
momentum components fulfill the usual canonical commutation relations
$[S_\alpha,P_\beta]=i\delta_{\alpha\beta}$. In the large-$N$ limit,
$N\to \infty$, the hard constraint $\mathbf{S}^2=N$ can
be replaced by one for the
thermodynamic average, $\langle \mathbf{S}^2 \rangle =N$, because fluctuations of the
magnitude of $\mathbf{S}$ are suppressed by the central limit theorem. The large-$N$ quantum
rotor is thus equivalent to the quantum spherical model of Ref.\ \onlinecite{Vojta96,*VojtaSchreiber96}
which is given by the Hamiltonian
\begin{equation}
H_S=\frac 1 2 P^2 + \frac 1 2 \omega_0^2 S^2 -h\,S +\mu(S^2-1)~.
\label{eq:H_S}
\end{equation}
Here, $S$ and $P$ represent the position and momentum of one rotor component, $\mu$ is a Lagrange multiplier
enforcing the constraint $\langle S^2 \rangle=1$, and $h$ is an external symmetry-breaking
field.\footnote{For the original rotor, this corresponds to a field coupling to all components,
$\mathbf{h} = h(1,1,\dots,1)$. This convention is convenient because the components remain equivalent
even in the presence of a field.}

We now couple (every component of) the rotor to a bath of harmonic
oscillators.\footnote{Equivalently, the $N$-component rotor is coupled to
$N$-component oscillators.}
In the conventional linear-coupling form,
the Hamiltonian describing the bath and its coupling to $S$ reads
\begin{equation}
H_B = \sum_j \left[\frac{p_j^2}{2m_j} + \frac{m_j}{2}\omega_j^2 q_j^2
     + \lambda_j q_j S + \frac{\lambda_j^2}{2m_j\omega_j^2}S^2\right],
\label{eq:H_B}
\end{equation}
with $q_j$, $p_j$, and $m_j$ being the position, momentum, and mass of the $j$-th oscillator.
The $\omega_j$ are the oscillator frequencies and $\lambda_j$ the coupling strengths between
the oscillators and $S$. The last term in the bracket is the usual counter term which insures
that the dissipation is invariant under translations in $S$.\cite{LCDFGZ87}
The coupling between the rotor and the bath is completely characterized by the spectral density
\begin{equation}
J(\omega) = \frac {\pi}{2} \sum_j \frac {\lambda_j^2}{m_j\omega_j}\delta(\omega-\omega_j)
\label{eq:J_definition}
\end{equation}
which we assume to be of power-law form
\begin{equation}
J(\omega) = 2\pi \bar\alpha \omega_c^{1-s} \omega^s, \qquad (0<\omega<\omega_c)~.
\label{eq:J_powerlaw}
\end{equation}
Here, $\bar\alpha$ is the dimensionless dissipation strength and $\omega_c$ is a cutoff frequency.
We will be interested mostly in the case of sub-Ohmic dissipation, $0<s<1$.

\section{Partition function and constraint equation}
\label{sec:constraint}
\subsection{Path integral formulation}
\label{subsec:path_integral}

We now derive a representation of the partition function in terms of an imaginary-time functional
integral. Because the sub-Ohmic rotor model
$H=H_S+H_B$ is equivalent to a system of coupled harmonic oscillators (with an additional self-consistency
condition), this can be done following
Feynman's path integral approach\cite{FeynmanHibbs_book65} with position and momentum eigenstates
as basis states. After integrating out the momentum variables, we arrive at the partition function
\begin{eqnarray}
Z  = \int D[S(\tau)]D[q_j(\tau)] ~e^{-{\cal A}_S -{\cal A}_B}~.
\label{eq:Z_SB}
\end{eqnarray}
The Euclidian action is given by
\begin{eqnarray}
{\cal A}_S &=& \int_0^{\beta} d\tau \left[ \frac 1 2 (\dot S^2 + \omega_0^2 S^2) - hS +\mu(S^2-1)  \right]\\
{\cal A}_B &=& \int_0^{\beta} d\tau \sum_j \left[\frac {m_j} 2 (\dot q_j^2 + \omega_j^2 q_j^2) +S \lambda_j q_j +\frac {\lambda_j^2 S^2}{2m_j\omega_j^2}\right ]
\label{eq:A_SB}
\end{eqnarray}
where the dot marks the derivative with respect to imaginary time $\tau$, and $\beta=1/T$ is the inverse temperature.

The bath action is quadratic in the $q_j$, we can thus exactly integrate out the bath modes. After a Fourier transformation
from imaginary time $\tau$ to Matsubara frequency $\omega_n$, this yields
$\int D[\tilde q_i(\omega_n)] \exp(-{\cal A}_B) = Z_B^0\exp(-{\cal A}_B{\prime})$
where $Z_B^0$ is the partition function of the unperturbed bath and
\begin{equation}
{\cal A}_B^{\prime} = T \sum_{\omega_n} \sum_j \frac {\lambda_j^2}{2 m_j} \, \frac{\omega_n^2}{\omega_j^2(\omega_n^2+\omega_j^2)} \, \tilde S(\omega_n) \tilde S(-\omega_n)~.
\label{eq:A_B'}
\end{equation}
The sum over $j$ can be turned into an integral over the spectral density $J(\omega)$. Carrying out this
integral gives
\begin{equation}
{\cal A}_B^{\prime} = \frac 1 2 T \sum_{\omega_n} \alpha \omega_c^{1-s} |\omega_n|^s  \tilde S(\omega_n) \tilde S(-\omega_n)
\label{eq:A_B'_integrated}
\end{equation}
with the dimensionless coupling constant $\alpha= 2 \pi \bar\alpha\, {\rm cosec}(\pi s/ 2)$.  Combining ${\cal A}_S$ and ${\cal A}_B^{\prime}$
yields the effective action of the sub-Ohmic rotor model as
\begin{eqnarray}
{\cal A}_{\rm eff} =&& -\beta\mu + \frac T 2 \sum_{\omega_n} \left(\epsilon + \alpha \omega_c^{1-s} |\omega_n|^s \right )
                     \tilde S(\omega_n) \tilde S(-\omega_n) \nonumber\\
                     &&-T\sum_{\omega_n} \tilde h(\omega_n) \tilde S(-\omega_n)~,
\label{eq:A_eff}
\end{eqnarray}
where $\epsilon=\omega_0^2+2\mu$
 is the renormalized distance from quantum criticality.
The $\omega_n^2$ term in ${\cal A}_S$ is subleading in the limit $\omega_n \to 0$.
It is thus irrelevant for the critical behavior at
the quantum critical point and has been dropped. The theory then needs a cutoff for
the Matsubara frequencies which we chose to be $\omega_c$.
Because the effective action is Gaussian, the partition function
$Z=Z_B^0\int D[\tilde S(\omega_n)] \exp(-{\cal A}_{\rm eff})$ is easily evaluated.
We find
\begin{eqnarray}
Z= && Z_B^0 \exp ({\beta\mu}) \prod_{\omega_n} \left[\frac {2 \pi}{T(\epsilon + \alpha \omega_c^{1-s}|\omega_n|^s)} \right]^{1/2} \times \nonumber\\
   && \times \exp\left[ \frac T 2 \sum_{\omega_n} \frac {\tilde h(\omega_n) \tilde h(-\omega_n)}{\epsilon + \alpha \omega_c^{1-s}|\omega_n|^s}  \right] ~.
\label{eq:Z}
\end{eqnarray}

\subsection{Solving the spherical constraint}
\label{subsec:constraint}

The spherical (large-$N$) constraint $\langle S^2 \rangle = 1$ can be easily derived from
the free energy $F=-T \ln Z$ by means of the relation
$0 = \partial F / \partial \mu$. In the case of a time-independent external field $h$
with Fourier components $\tilde h(\omega_n) = \delta_{n,0}h/T$, this yields
\begin{equation}
T\sum_{\omega_n} \frac 1 {\epsilon + \alpha \omega_c^{1-s}|\omega_n|^s} + \frac {h^2}{\epsilon^2} = 1~.
\label{eq:constraint}
\end{equation}
We now solve this equation, which gives the renormalized distance from criticality, $\epsilon$, as a function
of the external parameters $\alpha$, $T$, and $h$, in various limiting cases.

\subsubsection{$T=0$ and $h=0$}

At zero temperature, the sum over the Matsubara frequencies turns into an integral,
and the constraint equation reads
\begin{equation}
\frac 1 \pi \int_0^{\omega_c} d\omega  \frac {1}{\epsilon + \alpha \omega_c^{1-s}\omega^s}=1~.
\label{eq:constraintT0h0}
\end{equation}
For sub-Ohmic dissipation, $s<1$, a solution $\epsilon\ge0$ to this equation only exists for dissipation strengths
$\alpha$ below a critical value $\alpha_c$ because the integral converges at the lower bound
even for $\epsilon=0$. The value of $\alpha_c$ defines the location of the quantum critical point.
Performing the integral for $\epsilon=0$, we find $\alpha_c=1/[\pi(1-s)]$.
As we are interested in the critical behavior, we now solve the constraint equation for dissipation strengths
close to the critical one, $\alpha \lesssim \alpha_c$. We need to distinguish two cases: $1>s>1/2$ and
 $s<1/2$.

In the first case, the calculation can be performed by subtracting the
constraint equations at $\alpha$ and at $\alpha_c$ from each other. After moving the cutoff $\omega_c$ to
$\infty$, the resulting integral can be easily evaluated giving
\begin{equation}
\epsilon = \alpha \omega_c A^{s/(s-1)} (\alpha_c-\alpha)^{s/(1-s)} \qquad (s>1/2)~,
\label{eq:eps_T0h0_IFP}
\end{equation}
where $A=-(1/s)\, {\rm cosec}(\pi/s)$. In the case $s<1/2$, eq.\ (\ref{eq:constraintT0h0}) can be evaluated by a straight
Taylor expansion in $\alpha_c-\alpha$, resulting in
\begin{equation}
\epsilon = \alpha_c \omega_c B^{-1} (\alpha_c-\alpha) \qquad (s<1/2)~,
\label{eq:eps_T0h0_MFFP}
\end{equation}
with $B=1/[\pi(1-2s)]$. For $s<1/2$, the functional dependence of $\epsilon$ on $\alpha_c-\alpha$ thus becomes linear,
independent of $s$. As we will see later, this causes the transition to be of mean-field type.

For dissipation strengths above the critical value $\alpha_c$, the spherical constraint can only be solved by
\emph{not} transforming the sum over the Matsubara frequencies in (\ref{eq:constraint}) into the frequency integral in (\ref{eq:constraintT0h0}).
Instead, the $\omega_n=0$ Fourier component has to be treated separately.\footnote{This is analogous to the usual analysis
of Bose-Einstein condensation where the $\mathbf{q}=0$ mode has to be treated separately below the condensation
temperature.} Alternatively, one can explicitly introduce a nonzero average for one of the $N$ order parameter
components (see, e.g., Ref. \onlinecite{Sachdev_book99}). Both approaches are equivalent; we will follow the
first route in the next subsection.

\subsubsection{$T > 0$ and $h=0$}

At small but nonzero temperatures, an approximate solution of the spherical constraint (\ref{eq:constraint})
can be obtained by keeping the $\omega_n=0$ term in the frequency sum discrete while representing all other modes
in terms of an $\omega$-integral. This gives
\begin{equation}
\frac T \epsilon + \frac 1 \pi \int_0^{\omega_c} d\omega  \frac {1}{\epsilon + \alpha \omega_c^{1-s}\omega^s} = 1~.
\label{eq:constraintT<>0h0}
\end{equation}
We now solve this equation on the disordered side of the transition ($\alpha<\alpha_c$), at the critical dissipation strength $\alpha_c$,
and on the ordered side of the transition ($\alpha>\alpha_c$). We again need to distinguish the cases $1>s>1/2$ and $s<1/2$.

In the first case, we subtract the quantum critical ($T=0,h=0,\alpha=\alpha_c$) constraint from
(\ref{eq:constraintT<>0h0}). After evaluating the emerging integral, the following results are obtained
in the limit $T\to 0$ and $|\alpha-\alpha_c|$ small but fixed,
\begin{subequations}
\begin{eqnarray}
\epsilon &=& \frac \alpha {\alpha -\alpha_c} \, T \quad ~\qquad \qquad(\alpha > \alpha_c, ~s>1/2)~,\\
\epsilon &=& A^{-s} \alpha_c \omega_c^{1-s} T^s \qquad \quad ~ (\alpha = \alpha_c,~ s>1/2)~,\\
\epsilon &=& \epsilon_0 + \frac \alpha {\alpha_c-\alpha}\, \frac  s {1-s} T \quad (\alpha < \alpha_c,~ s>1/2)~.
\end{eqnarray}
\label{eq:eps_T<>h0_IFP}
\end{subequations}
Here, $\epsilon_0$ is the zero-temperature value given in (\ref{eq:eps_T0h0_IFP}) and $A=-(1/s)\, {\rm cosec}(\pi/s)$
as above. For $s<1/2$, we expand (\ref{eq:constraintT<>0h0}) in $\alpha-\alpha_c$ and find
\begin{subequations}
\begin{eqnarray}
\epsilon &=& \frac \alpha {\alpha -\alpha_c}\, T \quad ~\qquad \qquad(\alpha > \alpha_c, ~s<1/2)~,\\
\epsilon &=& B^{-1/2} \alpha_c \omega_c^{1/2} T^{1/2} \qquad  (\alpha = \alpha_c,~ s<1/2)~,\\
\epsilon &=& \epsilon_0 + \frac \alpha {\alpha_c-\alpha}\, T \qquad \quad ~(\alpha < \alpha_c,~ s<1/2)~,
\end{eqnarray}
\label{eq:eps_T<>h0_MFFP}
\end{subequations}
with $\epsilon_0$ given in (\ref{eq:eps_T0h0_MFFP}) and $B=1/[\pi(1-2s)]$ as above.

\subsubsection{$T = 0$ and $h \ne 0$}

At zero temperature, but in the presence of an external field, the spherical constraint reads
\begin{equation}
\frac 1 \pi \int_0^{\omega_c} d\omega  \frac {1}{\epsilon + \alpha \omega_c^{1-s}\omega^s} + \frac {h^2}{\epsilon^2}= 1~.
\label{eq:constraintT0h<>0}
\end{equation}
Proceeding in analogy to the last subsection, we determine the distance $\epsilon$ from criticality
in the limit $h\to 0$ and $|\alpha-\alpha_c|$ small but fixed. In the case $1>s>1/2$,
we obtain
\begin{subequations}
\begin{eqnarray}
\epsilon &=& \left(\frac \alpha {\alpha -\alpha_c}\right)^{1/2} h \qquad \qquad ~(\alpha > \alpha_c, ~s>1/2)~,~~\\
\epsilon &=& \left( A^{-s} \alpha_c \omega_c^{1-s} h^{2s} \right)^{1/(s+1)} ~~ (\alpha = \alpha_c,~ s>1/2)~,~~\\
\epsilon &=& \epsilon_0 + \frac \alpha {\alpha_c-\alpha}\, \frac  s {1-s} \, \frac {h^2} {\epsilon_0} \quad~~ (\alpha < \alpha_c,~ s>1/2)~,~~
\end{eqnarray}
\label{eq:eps_T0h<>0_IFP}
\end{subequations}
where $\epsilon_0$ is the zero-field value given in (\ref{eq:eps_T0h0_IFP}) and $A=-(1/s)\, {\rm cosec}(\pi/s)$
as above. For $s<1/2$, the corresponding results read
\begin{subequations}
\begin{eqnarray}
\epsilon &=& \left(\frac \alpha {\alpha -\alpha_c}\right)^{1/2} h \qquad \quad (\alpha > \alpha_c, ~s<1/2)~,~\\
\epsilon &=& \left( B^{-1} \alpha_c^2 \omega_c h^2 \right)^{1/3} \qquad (\alpha = \alpha_c,~ s<1/2)~,~\\
\epsilon &=& \epsilon_0 + \frac \alpha {\alpha_c-\alpha}\, \frac {h^2} {\epsilon_0} \qquad \quad (\alpha < \alpha_c,~ s<1/2)~,
\end{eqnarray}
\label{eq:eps_T0h<>0_MFFP}
\end{subequations}
with $\epsilon_0$ given in (\ref{eq:eps_T0h0_MFFP}) and $B=1/[\pi(1-2s)]$ as above.

\section{Observables at the quantum phase transition}
\label{sec:observables}

After having solved the spherical constraint, we now turn to the behavior of observables
at the quantum critical point.

\subsection{Magnetization}

The magnetization $M=\langle S \rangle$  follows from (\ref{eq:Z})
via $M=-\partial F / \partial h = T \partial(\ln Z)/\partial h$. This simply gives
\begin{equation}
M=h / \epsilon~.
\label{eq:M}
\end{equation}
To find the zero-temperature spontaneous magnetization in the ordered phase, we need evaluate (\ref{eq:M}) for
$T=0$, $\alpha>\alpha_c$, and $h \to 0$. Using equations (\ref{eq:eps_T0h<>0_IFP}a) and (\ref{eq:eps_T0h<>0_MFFP}a), we find
\begin{equation}
M= \sqrt{(\alpha-\alpha_c)/\alpha}
\label{eq:M_spon}
\end{equation}
for the entire range $1>s>0$. The order parameter exponent $\beta$ thus takes the value 1/2 in the
entire $s$-range. For $T>0$, $\epsilon$ does not vanish even in the limit $h \to 0$.
The spontaneous magnetization is therefore identical to zero for any nonzero temperature,
independent of the dissipation strength $\alpha$.

The critical magnetization-field curve of the quantum phase transition
can be determined by analyzing (\ref{eq:M}) for $T=0$, $\alpha=\alpha_c$, and
nonzero $h$. In the case $1>s>1/2$, inserting (\ref{eq:eps_T0h<>0_IFP}b) into (\ref{eq:M}) yields
\begin{equation}
M= \left( A^s \alpha_c^{-1} \omega_c^{-(1-s)} h^{1-s} \right)^{1/(1+s)} \qquad (s>1/2)
\label{eq:m-H_IFP}
\end{equation}
which implies a critical exponent $\delta=(1+s)/(1-s)$. For $s<1/2$, we instead get the relation
\begin{equation}
M= \left ( B \alpha_c^{-2} \omega_c^{-1} h \right)^{1/3} \qquad (s<1/2)~.
\label{eq:m-H_MFFP}
\end{equation}
The critical exponent $\delta$ thus takes the mean-field value of 3.

\subsection{Susceptibility}

The Matsubara susceptibility can be calculated by taking the second derivative of
$\ln Z$ in (\ref{eq:Z}) with respect to the Fourier components of the field, yielding
\begin{equation}
\chi({i\omega_n}) = \frac 1 {\epsilon + \alpha \omega_c^{1-s}|\omega_n|^s}~.
\label{eq:chi(omega_n)}
\end{equation}
We first discuss the static susceptibility $\chi_{\rm st}=\chi(0) = 1/\epsilon$ in the case
$1>s>1/2$. To find the zero-temperature, zero-field susceptibility in the
disordered (delocalized) phase, $\alpha < \alpha_c$, we use (\ref{eq:eps_T0h0_IFP}) for $\epsilon$,
which results in
\begin{equation}
\chi_{\rm st} = \alpha^{-1}\omega_c^{-1} A^{s/(1-s)} (\alpha_c-\alpha)^{-s/(1-s)} \quad(s>1/2).
\label{chi_st_T0H0_IFP}
\end{equation}
The susceptibility exponent thus takes the value $\gamma=s/(1-s)$.

For dissipation strengths $\alpha \ge \alpha_c$, the susceptibility diverges in the limit $T \to 0$.
The temperature dependencies follow from substituting
(\ref{eq:eps_T<>h0_IFP}a) and (\ref{eq:eps_T<>h0_IFP}b) into $\chi_{\rm st}= 1/\epsilon$. This
yields
\begin{subequations}
\begin{eqnarray}
\chi_{\rm st} &=& \frac{\alpha-\alpha_c} {\alpha} \, T^{-1} \qquad ~ (\alpha > \alpha_c,~ s>1/2)~,~\\
\chi_{\rm st} &=& \omega_c^{s-1}\alpha_c^{-1}A^s T^{-s}  \quad (\alpha = \alpha_c,~ s>1/2)~.~
\end{eqnarray}
\label{eq:ch_st_T<>0H0_IFP}
\end{subequations}
In the ordered (localized) phase, we thus find Curie behavior with an effective moment of
$M^2=(\alpha-\alpha_c)/\alpha$  in agreement with (\ref{eq:M_spon}).

The static susceptibility in the case $s<1/2$ is obtained analogously. Using (\ref{eq:eps_T0h0_MFFP}),
the zero-temperature, zero-field susceptibility reads
\begin{equation}
\chi_{\rm st} = \alpha^{-1}\omega_c^{-1} B (\alpha_c-\alpha)^{-1} \qquad(s<1/2)~,
\label{chi_st_T0H0_MFFP}
\end{equation}
implying that the susceptibility exponent takes the mean-field value $\gamma=1$. From (\ref{eq:eps_T<>h0_MFFP}b),
we obtain the temperature dependence of $\chi_{\rm st}$ at the critical damping strength,
\begin{equation}
\chi_{\rm st} = \omega_c^{-1}\alpha_c^{-1/2} B^{1/2} T^{-1/2}  \quad (\alpha = \alpha_c,~ s<1/2)~.~
\label{{eq:ch_st_T<>0H0_MFFP}}
\end{equation}
In the ordered phase, the behavior for $s<1/2$ is identical to that for $s>1/2$ given in (\ref{eq:ch_st_T<>0H0_IFP}a).

We now turn to the dynamic susceptibility. To compute the retarded susceptibility $\chi(\omega)$, we need to
analytically continue the Matsubara susceptibility by performing a Wick rotation to real frequencies,
$i\omega_n \to \omega+i0$. A direct transformation of (\ref{eq:chi(omega_n)}) is hampered by the
non-analytic frequency dependence $|\omega_n|^s$. We therefore go back to a representation of
the dynamic term in the susceptibility in terms of discrete bath modes [see the action (\ref{eq:A_B'})].
As this representation is analytic in $\omega_n$, the Wick rotation can be performed easily. We then  carry out
the integration over the spectral density \emph{after} the Wick rotation. The resulting dynamical susceptibility reads
\begin{equation}
\chi(\omega) = \frac 1 {\epsilon + \alpha \omega_c^{1-s} |\omega|^s \left[ \cos(\pi s/2) -i \sin(\pi s/2)\textrm{sgn}(\omega)   \right]}~.
\label{eq:chi(omega)}
\end{equation}
At quantum criticality ($\alpha=\alpha_c$, $T=0$, $h=0$), the real and imaginary parts of the dynamic susceptibility
simplify to
\begin{eqnarray}
\textrm{Re} \chi(\omega) = \frac {\cos(\pi s/2)}{\alpha_c\omega_c^{1-s}|\omega|^s}\,,~
\textrm{Im} \chi(\omega) = \frac {\sin(\pi s/2)\textrm{sgn}(\omega)}{\alpha_c\omega_c^{1-s}|\omega|^s} \quad~
\label{eq:ReImchi}
\end{eqnarray}
in the entire range $1>s>0$. Comparing this with the temperature dependencies (\ref{eq:ch_st_T<>0H0_IFP}b) and
(\ref{{eq:ch_st_T<>0H0_MFFP}}), we note that the results for $s<1/2$ violate $\omega/T$ scaling while those
for $1>s>1/2$ are compatible with it.

\subsection{Correlation time}

To find the inverse correlation time (characteristic energy) $\Delta=\xi_t^{-1}$,
we parameterize the inverse susceptibility as
$\epsilon +\alpha\omega_c^{1-s} |\omega_n|^s = \epsilon (1+ |\omega_n/\Delta|^s)$.
This implies the relation
\begin{equation}
\Delta = \left (\epsilon \alpha^{-1}\omega_c^{s-1}  \right)^{1/s}~.
\label{eq:Delta_def}
\end{equation}
The dependence of the inverse correlation time on the tuning parameter $\alpha$
at zero temperature and field in the case $1>s>1/2$ is obtained by inserting
(\ref{eq:eps_T0h0_IFP}) into (\ref{eq:Delta_def}). In the disordered phase,
$\alpha<\alpha_c$, this gives
\begin{equation}
\Delta = \omega_c A^{-1/(1-s)} (\alpha_c-\alpha)^{1/(1-s)} \qquad (s>1/2)~.
\label{eq:Delta_T0H0_IFP}
\end{equation}
The correlation-time critical exponent therefore reads $\nu z = 1/(1-s)$. Note that this exponent
is sometimes called just $\nu$ rather than $\nu z$ in the literature
on impurity transitions.
We follow the general convention for quantum phase transitions where $\nu$ describes the divergence
of the correlation \emph{length} while $\nu z$ that of the correlation \emph{time}. By substituting
(\ref{eq:eps_T<>h0_IFP}b) into (\ref{eq:Delta_def}),
we can also determine the dependence of $\Delta$ on temperature at $\alpha=\alpha_c$
and $h=0$. We find $\Delta =A^{-1} T$. The characteristic energy thus scales with $T$,
as expected from naive scaling.

In the case $s<1/2$, the zero-temperature, zero-field correlation time in the disordered
phase behaves as [using (\ref{eq:eps_T0h0_MFFP})]
\begin{equation}
\Delta = \omega_c B^{-1/s} (\alpha_c-\alpha)^{1/s} \qquad (s<1/2)~,
\label{eq:Delta_T0H0_MFFP}
\end{equation}
resulting in the mean-field value $\nu z = 1/s$  for the correlation time critical exponent.
The dependence of $\Delta$ on temperature at $\alpha=\alpha_c$
and $h=0$ follows from (\ref{eq:eps_T<>h0_MFFP}b); it reads $\Delta = B^{-1/(2s)} \omega_c^{(2s-1)/(2s)} T^{1/(2s)}$.
The characteristic energy thus scales differently than the temperature, in disagreement
with naive scaling.

\subsection{Scaling form of the equation of state}

A scaling form of the equation of state for $1>s>1/2$ can be determined by subtracting the quantum critical
($T=0$, $h=0$, $\alpha=\alpha_c$) spherical constraint from the general constraint
(\ref{eq:constraint}). After performing the resulting integral, we find
\begin{equation}
\frac {\alpha_c-\alpha}{\alpha} + \frac {h^2}{\epsilon^2} + \frac T \epsilon = A\epsilon^{-1+1/s} \alpha^{-1/s} \omega_c^{1-1/s}~.
\label{eq:EOS}
\end{equation}
We substitute $\epsilon =h/M$ [from (\ref{eq:M})]; and after some lengthy but
straight forward algebra, this equation can be written in the scaling form
\begin{equation}
X\left(M/r^{1/2},h/r^{(1+s)/(2-2s)}, T/r^{1/(1-s)} \right) = 0,
\label{eq:EOS_scaling}
\end{equation}
with $X$ being the scaling function, and $r=(\alpha-\alpha_c)/\alpha$ being the reduced
distance from criticality. This scaling form can be used to reproduce the
critical exponents $\beta=1/2$, $\gamma=s/(1-s)$, and $\delta=(1+s)/(1-s)$ found above.

For $s<1/2$, the same approach gives a scaling equation containing the mean-field
exponents $\beta=1/2$, $\gamma=1$, and $\delta=3$. Moreover, an explicit dependence
on the cutoff for the Matsubara frequencies remains.

\subsection{Entropy and specific heat}

Within our path integral approach, thermal properties are somewhat harder to calculate
than magnetic properties because the measure of the path integral explicitly depends
on temperature. As the spherical model is equivalent to a set of coupled harmonic
oscillators, we can use the ``remarkable formulas'' derived by Ford et
al.,\cite{FordLewisOConnell85,*FordLewisOConnell88} which express the free energy
(and internal energy) of a quantum oscillator in a heat bath in terms of its susceptibility and the free
energy (and internal energy) of a free oscillator. For our spherical model, they read
\begin{eqnarray}
F_S &=& -\mu + \frac 1 \pi \int_0^\infty d\omega \, F_f(\omega,T)\, \textrm{Im} \left[ \frac d {d\omega} \ln\chi(\omega) \right]~,
\label{eq:remarkable_F}\\
U_S &=& -\mu + \frac 1 \pi \int_0^\infty d\omega \, U_f(\omega,T)\, \textrm{Im} \left[ \frac d {d\omega} \ln\chi(\omega) \right]~.
\label{eq:remarkable_U}
\end{eqnarray}
Here, $F_f(\omega, T) = T\ln [2 \sinh(\omega/2T)]$ and $U_f(\omega, T) = (\omega/2) \coth(\omega/2T)$. The extra $-\mu$ terms stem
from the spherical constraint. Note that the free energy in (\ref{eq:remarkable_F}) is the difference between the free energy of
the coupled rotor-bath system and that of the unperturbed bath, $F_S = F - F_B^0= -T\ln(Z/Z_B^0)$. The same holds true for
the internal energy, $U_S =U - U_B^0$.

The frequency derivative of $\ln\chi(\omega)$ can be calculated from (\ref{eq:chi(omega)}), giving
\begin{eqnarray}
\label{eq:dlnchidomega}
&&\textrm{Im} \left[ \frac d {d\omega} \ln\chi(\omega) \right] = \\
&&~=\frac {\epsilon s \alpha \omega_c^{1-s} \omega^{s-1} \sin(\pi s/2)}
  {[\epsilon + \alpha \omega_c^{1-s} \omega^{s} \cos(\pi s/2)]^2+[\alpha \omega_c^{1-s} \omega^{s} \sin(\pi s/2)]^2}.
\nonumber
\end{eqnarray}

To calculate the impurity entropy  $S_S=(U_S-F_S)/T$, we insert (\ref{eq:dlnchidomega}) into (\ref{eq:remarkable_F}) and (\ref{eq:remarkable_U})
and perform the resulting integral. In the disordered phase, $\alpha<\alpha_c$, the entropy behaves as
\begin{equation}
S_S = D \, \alpha \, \omega_c^{1-s} T^s / \epsilon_0
\label{eq:S_disordered}
\end{equation}
in the limit $T\to 0$ for all $s$ in the sub-Ohmic range $1>s>0$.
Here, $\epsilon_0$ is the zero-temperature renormalized distance from
criticality given in (\ref{eq:eps_T0h0_IFP}), and $D$ is an $s$-dependent constant.
Upon approaching criticality, $\alpha \to \alpha_c$, the prefactor of the $T^s$ power-law
diverges, suggesting a weaker temperature dependence at criticality.
The specific heat can be calculated from $C_S=T(\partial S_S / \partial T)$, it thus behaves
as $Ds \alpha  \omega_c^{1-s} T^s / \epsilon_0$.

We now turn to the critical dissipation strength, $\alpha=\alpha_c$, For $1>s>1/2$,
we find a temperature-independent but
non-universal ($s$-dependent) entropy in the limit of low temperatures. For $s<1/2$, the
impurity entropy diverges logarithmically as $\ln(\omega_0/T)$ with $T \to 0$.
In the ordered phase, $\alpha>\alpha_c$, we find a logarithmically diverging entropy
for all $s$ between 0 and 1.

At first glance, these logarithmic divergencies appear to violate the third law of thermodynamics.
We emphasize, however, that the impurity entropy represents the
difference between the entropy of the coupled rotor-bath system and that of the unperturbed bath.
Because the bath is infinite, the entropy thus involves an \emph{infinite} number of degrees of freedom
and does not have to remain finite. Whether the logarithmic divergence occurs
only in the large-$N$ limit or also for finite-$N$ rotors remains a question for
the future.

We note in passing that the entropy of classical spherical models\cite{BerlinKac52,Joyce66}
also diverges in the limit $T\to 0$ (even when measured per degree of freedom).
In these models, the diverges occurs because the classical description becomes invalid
at sufficiently low temperatures. It can be cured by going from the classical spherical model
to the quantum spherical  model.\cite{Vojta96} This implies that the diverging entropy in
the ordered phase of the sub-Ohmic rotor model is caused by a different mechanism than that
in the classical spherical model.

\section{Conclusions}
\label{sec:conclusions}

\begin{table}[tb!]
\renewcommand{\arraystretch}{1.2}
\begin{tabular*}{70mm}{c@{\extracolsep\fill}cc}
\hline\hline
            &  $1>s>1/2$  & $s<1/2$    \\
\hline
$\beta$     &  $1/2$        & $1/2$      \\
$\gamma$    &  $s/(1-s)$    & $1$        \\
$\delta$    & $(1+s)/(1-s)$ & $3$        \\
$\nu z$     & $1/(1-s)$     & $1/s$      \\
$\eta$      & $2-s$         & $2-s$      \\
\hline\hline
\end{tabular*}
\caption{Critical exponents of the sub-Ohmic quantum rotor model.}
\label{table:exponents}
\end{table}

In summary, we have investigated the quantum critical behavior of a large-$N$ quantum
rotor coupled to a sub-Ohmic bosonic bath characterized by a power-law spectral density
$J(\omega) \sim \omega^s$ with $0<s<1$. As this model can be solved exactly, it provides
a reliable reference point for the discussion of more complex and realistic impurity quantum phase
transitions.
We find that all critical exponents take their mean-field values if the bath exponent $s$ is below $1/2$.
In contrast, for $1>s>1/2$, the exponents display nontrivial, $s$-dependent
values.
A summary of the exponent values in both cases in shown in table
\ref{table:exponents}.
The exponent $\eta$ sticks to its mean-field value $2-s$ in the entire
region $1>s>0$, in agreement with renormalization group arguments on the absence
of field renormalization for long-range interactions.\cite{FisherMaNickel72,*Suzuki73a,*Suzuki73b}
The fact that the order parameter exponent $\beta$ is 1/2 in the entire range $1>s>0$ is a results
of the large-$N$ limit; it generically takes this value in spherical models.

Moreover, the behaviors of the dynamic susceptibility and inverse correlation time are
compatible with $\omega/T$ scaling for $1>s>1/2$, while they violate $\omega/T$ scaling
for $s<1/2$. We conclude that the quantum phase transition of the sub-Ohmic quantum rotor model
is controlled by an interacting renormalization group fixed point in the case $1>s>1/2$.
In contrast, the transition is controlled by a noninteracting (Gaussian) fixed point
for $s<1/2$.

We now turn to the question of the quantum-to-classical mapping. The classical counterpart
of the sub-Ohmic quantum rotor model is a one-dimensional classical Heisenberg chain with
long-range interactions that decay as $1/r^{1+s}$ with distance $r$.
The spherical (large-$N$) version of this model was solved
by Joyce;\cite{Joyce66} its critical exponents  are identical to
that of the sub-Ohmic quantum rotor found here. The
quantum-to-classical mapping is thus valid.

The properties of our quantum \emph{rotor} model must be contrasted with the behavior of
the Bose-Kondo model which describes a continuous
symmetry quantum \emph{spin} coupled to a bosonic bath. For this system, the
quantum-to-classical mapping appears to be inapplicable.\cite{SachdevBuragohainVojta99}
A related observation has been made in a Bose-Fermi-Kondo model.\cite{ZKSG04}
The main difference between a rotor and a quantum spin is the presence
of the Berry phase term in the action of the latter.
Our results thus support the conjecture that this Berry phase term,
which is complex and has no classical analog, causes the inapplicability
of the quantum-to-classical mapping.

\section*{Acknowledgements}

We acknowledge helpful discussions with M. Vojta and S. Kirchner.
This work has been supported in part by the NSF under grant no. DMR-0906566.

\bibliographystyle{apsrev4-1}
\bibliography{../00Bibtex/rareregions}
\end{document}